\documentstyle[twoside,psfig]{article}

\catcode`\@=11
\long\def\@makefntext#1{
\protect\noindent \hbox to 3.2pt {\hskip-.9pt  
$^{{\eightrm\@thefnmark}}$\hfil}#1\hfill}		%CAN BE USED 

\def\@makefnmark{\hbox to 0pt{$^{\@thefnmark}$\hss}}	%ORIGINAL 
	
\def\ps@myheadings{\let\@mkboth\@gobbletwo
\def\@oddhead{\hbox{}
\rightmark\hfil\eightrm\thepage}   
\def\@oddfoot{}\def\@evenhead{\eightrm\thepage\hfil
\leftmark\hbox{}}\def\@evenfoot{}
\def\sectionmark##1{}\def\subsectionmark##1{}}

\oddsidemargin=\evensidemargin
\newcounter{sectionc}\newcounter{subsectionc}\newcounter{subsubsectionc}
\renewcommand{\section}[1] {\vspace{12pt}\addtocounter{sectionc}{1} 
\setcounter{subsectionc}{0}\setcounter{subsubsectionc}{0}\noindent 
	{\tenbf\thesectionc. #1}\par\vspace{5pt}}
\renewcommand{\subsection}[1] {\vspace{12pt}\addtocounter{subsectionc}{1} 
	\setcounter{subsubsectionc}{0}\noindent 
	{\bf\thesectionc.\thesubsectionc. {\kern1pt \bfit #1}}\par\vspace{5pt}}
\renewcommand{\subsubsection}[1] {\vspace{12pt}\addtocounter{subsubsectionc}{1}
	\noindent{\tenrm\thesectionc.\thesubsectionc.\thesubsubsectionc.
	{\kern1pt \tenit #1}}\par\vspace{5pt}}
\newcommand{\nonumsection}[1] {\vspace{12pt}\noindent{\tenbf #1}
	\par\vspace{5pt}}

\newcounter{appendixc}
\newcounter{subappendixc}[appendixc]
\newcounter{subsubappendixc}[subappendixc]
\renewcommand{\thesubappendixc}{\Alph{appendixc}.\arabic{subappendixc}}
\renewcommand{\thesubsubappendixc}
	{\Alph{appendixc}.\arabic{subappendixc}.\arabic{subsubappendixc}}

\renewcommand{\appendix}[1] {\vspace{12pt}
        \refstepcounter{appendixc}
        \setcounter{figure}{0}
        \setcounter{table}{0}
        \setcounter{lemma}{0}
        \setcounter{theorem}{0}
        \setcounter{corollary}{0}
        \setcounter{definition}{0}
        \setcounter{equation}{0}
        \renewcommand{\thefigure}{\Alph{appendixc}.\arabic{figure}}
        \renewcommand{\thetable}{\Alph{appendixc}.\arabic{table}}
        \renewcommand{\theappendixc}{\Alph{appendixc}}
        \renewcommand{\thelemma}{\Alph{appendixc}.\arabic{lemma}}
        \renewcommand{\thetheorem}{\Alph{appendixc}.\arabic{theorem}}
        \renewcommand{\thedefinition}{\Alph{appendixc}.\arabic{definition}}
        \renewcommand{\thecorollary}{\Alph{appendixc}.\arabic{corollary}}
        \renewcommand{\theequation}{\Alph{appendixc}.\arabic{equation}}
        \noindent{\tenbf Appendix \theappendixc #1}\par\vspace{5pt}}
\newcommand{\subappendix}[1] {\vspace{12pt}
        \refstepcounter{subappendixc}
        \noindent{\bf Appendix \thesubappendixc. {\kern1pt \bfit #1}}
	\par\vspace{5pt}}
\newcommand{\subsubappendix}[1] {\vspace{12pt}
        \refstepcounter{subsubappendixc}
        \noindent{\rm Appendix \thesubsubappendixc. {\kern1pt \tenit #1}}
	\par\vspace{5pt}}

\newcommand{\textlineskip}{\baselineskip=13pt}
\newcommand{\smalllineskip}{\baselineskip=10pt}

\def\eightcirc{
\begin{picture}(0,0)
\put(4.4,1.8){\circle{6.5}}
\end{picture}}
\def\eightcopyright{\eightcirc\kern2.7pt\hbox{\eightrm c}} 

\newcommand{\copyrightheading}[1]
	{\vspace*{-2.5cm}\smalllineskip{\flushleft
	{\footnotesize International Journal of Modern Physics D #1}\\
	{\footnotesize $\eightcopyright$\, World Scientific Publishing
	 Company}\\
	 }}

\newcommand{\publisher}[2]{{\begin{center}\footnotesize\smalllineskip 
	Received #1\\
	Revised #2
	\end{center}
	}}

\def\abstracts#1#2#3{{
	\centering{\begin{minipage}{4.5in}\footnotesize\baselineskip=10pt
	\parindent=0pt #1\par 
	\parindent=15pt #2\par
	\parindent=15pt #3
	\end{minipage}}\par}} 

\renewenvironment{thebibliography}[1]
	{\frenchspacing
	 \ninerm\baselineskip=11pt
	 \begin{list}{\arabic{enumi}.}
	{\usecounter{enumi}\setlength{\parsep}{0pt}
	 \setlength{\leftmargin 12.7pt}{\rightmargin 0pt} %FOR 1--9 ITEMS
	 \setlength{\itemsep}{0pt} \settowidth
	{\labelwidth}{#1.}\sloppy}}{\end{list}}

\newcounter{itemlistc}
\newcounter{romanlistc}
\newcounter{alphlistc}
\newcounter{arabiclistc}

\newcommand{\fcaption}[1]{
        \refstepcounter{figure}
        \setbox\@tempboxa = \hbox{\footnotesize Fig.~\thefigure. #1}
        \ifdim \wd\@tempboxa > 5in
           {\begin{center}
        \parbox{5in}{\footnotesize\smalllineskip Fig.~\thefigure. #1}
            \end{center}}
        \else
             {\begin{center}
             {\footnotesize Fig.~\thefigure. #1}
              \end{center}}
        \fi}

\newcommand{\tcaption}[1]{
        \refstepcounter{table}
        \setbox\@tempboxa = \hbox{\footnotesize Table~\thetable. #1}
        \ifdim \wd\@tempboxa > 5in
           {\begin{center}
        \parbox{5in}{\footnotesize\smalllineskip Table~\thetable. #1}
            \end{center}}
        \else
             {\begin{center}
             {\footnotesize Table~\thetable. #1}
              \end{center}}
        \fi}

\def\@citex[#1]#2{\if@filesw\immediate\write\@auxout
	{\string\citation{#2}}\fi
\def\@citea{}\@cite{\@for\@citeb:=#2\do
	{\@citea\def\@citea{,}\@ifundefined
	{b@\@citeb}{{\bf ?}\@warning
	{Citation `\@citeb' on page \thepage \space undefined}}
	{\csname b@\@citeb\endcsname}}}{#1}}

\newif\if@cghi
\def\cite{\@cghitrue\@ifnextchar [{\@tempswatrue
	\@citex}{\@tempswafalse\@citex[]}}
\def\citelow{\@cghifalse\@ifnextchar [{\@tempswatrue
	\@citex}{\@tempswafalse\@citex[]}}
\def\@cite#1#2{{$\null^{#1}$\if@tempswa\typeout
	{IJCGA warning: optional citation argument 
	ignored: `#2'} \fi}}

\def\pmb#1{\setbox0=\hbox{#1}
	\kern-.025em\copy0\kern-\wd0
	\kern.05em\copy0\kern-\wd0
	\kern-.025em\raise.0433em\box0}

\def\fnt#1#2{\footnotetext{\kern-.3em
	{$^{\mbox{\scriptsize #1}}$}{#2}}}

\def\@makefnmark{\hbox to 0pt{$^{\@thefnmark}$\hss}}	%ORIGINAL 
	
\def\ps@myheadings{%
    \let\@oddfoot\@empty\let\@evenfoot\@empty
    \def\@evenhead{\slshape\leftmark\hfil}%       %EVEN PAGE
    \def\@oddhead{\hfil{\slshape\rightmark}}%     %ODD PAGE
    \let\@mkboth\@gobbletwo
    \let\sectionmark\@gobble
    \let\subsectionmark\@gobble
    }
\font\tenrm=cmr10
\font\tenit=cmti10 
\font\tenbf=cmbx10
\font\bfit=cmbxti10 at 10pt
\font\ninerm=cmr9

\font\eightrm=cmr8

\def\qed{\hbox{${\vcenter{\vbox{			%HOLLOW SQUARE
   \hrule height 0.4pt\hbox{\vrule width 0.4pt height 6pt
   \kern5pt\vrule width 0.4pt}\hrule height 0.4pt}}}$}}

\def\lsim{\lower.5ex\hbox{$\; \buildrel < \over \sim \;$}}
\def\gsim{\lower.5ex\hbox{$\; \buildrel > \over \sim \;$}}
\begin{document}
\setlength{\textheight}{7.7truein}  %for 2nd page onwards

\thispagestyle{empty}

\markboth{\protect{\footnotesize\it Pseudo-Schwarzschild description 
of accretion-powered spherical outflow
}}{\protect{\footnotesize\it Pseudo-Schwarzschild description 
of accretion-powered spherical outflow
}}

\normalsize\textlineskip

\setcounter{page}{1}

\copyrightheading{}	%{Vol.~0, No.~0 (2000) 000--000}

\vspace*{0.88truein}

%\fpage{1}
\centerline{\bf PSEUDO-SCHWARZSCHILD DESCRIPTION OF}
\vspace*{0.035truein}
\centerline{\bf ACCRETION-POWERED SPHERICAL OUTFLOW}
%\footnote{For
%the title, try not to use more than 3 lines. Typeset the title
%in 10 pt Times Roman, uppercase and boldface.}}
\vspace*{0.37truein}

\centerline{\footnotesize TAPAS K.  DAS\footnote{\it tapas@iucaa.ernet.in}}
%\footnote{Typeset names in
%10 pt Times Roman, uppercase. Use the footnote to indicate the
%present or permanent address of the author.}}
\baselineskip=12pt
\centerline{\footnotesize\it Inter University centre for Astronomy and 
Astrophysics}
\baselineskip=10pt
\centerline{\footnotesize\it Post Bag 4 Ganeshkhind Pune 411 007 INDIA}
%\footnote{State completely without abbreviations, the
%affiliation and mailing address, including country. Typeset in 8
%pt Times Italic.}}
\vspace*{10pt}
\centerline{\footnotesize Present Address}
\baselineskip=12pt
\centerline{\footnotesize\it Division of Astronomy, University
of California at Los Angeles (UCLA)}
\baselineskip=10pt
\centerline{\footnotesize\it Box 951562, Los Angeles, CA 90095-1562, USA.}
%\vspace*{0.225truein}
\publisher{(received date)}{(revised date)}

\vspace*{0.21truein}
\abstracts{
Using two different pseudo-Schwarzschild potentials proposed by Artemova
et. al,$^{1}$ we formulate and solve the equations governing spherically
symmetric transonic inflow and outflow in presence of a relativistic
hadronic pressure mediated steady, standing, spherical shock around the
central compact object and then we self-consistently connect the
accretion-wind solutions to calculate the mass outflow rate $R_{\dot m}$
in terms of minimum number of flow parameters. Also we study the dependence
of this rate on various boundary conditions governing the flow.
}{}{}

%\textlineskip			%) USE THIS MEASUREMENT WHEN THERE IS
%\vspace*{12pt}			%) NO SECTION HEADING

\vspace*{1pt}\textlineskip	%) USE THIS MEASUREMENT WHEN THERE IS
\vskip 0.5cm
\noindent
\hrule
\noindent
{\bf Published in International Journal of Modern Physics D, 2002,
Volume 11, Issue 08, pp. 1285-1303.}
\hrule
\vskip 0.25cm
\section{Introduction}	%) A SECTION HEADING
\noindent
For spherically symmetric accretion onto a non-rotating compact object, a novel
mechanism was proposed by Protheros and Kazanas,$^{2}$
(PK83 hereafter) and Kazanas and Ellison$^{3}$ (KE86 hereafter),
in which the kinetic energy of the
accreting material 
was shown to be randomized by incorporating a steady, collision-less, 
relativistic hadronic-pressure-supported spherical shock surface around the
accreting Schwarzschild black hole which produces a non-thermal spectrum of 
relativistic protons.
Recently Das$^{4}$ (D99 hereafter) has
explicitly calculated the exact location (radial
distance measured from the
central accretor in units of Schwarzschild radius $r_g=\frac{2GM_{BH}}{c^2}$,
where $M_{BH}$ is the mass of the black hole, $G$ is the Universal gravitational
constant and $c$ is the velocity of light in vacuum)
of the above mentioned shock
in terms of only three accretion parameters, namely, the specific energy of
the flow ${\cal E}$, 
accretion rate ${\dot M}_{Edd}$
(scaled in units of Eddington rate) and the adiabatic
index $\gamma_{in}$ of the inflow 
for spherically-symmetric, transonic accretion of
adiabatic fluid onto a Schwarzschild black hole.
By solving the set of hydrodynamic equations describing the
motion of accreting material under the influence of modified Newtonian 
potential
proposed by Paczy\'nski and Wiita,$^5$ it has been shown there in D99
that it is possible to construct a self-consistent inflow-outflow system
where a part of the accreting material may be blown as wind from the
spherical shock surface and the mass outflow rate $R_{\dot m}$ (the measure of the
fraction of accreting material being `kicked off' as wind) was computed
in terms of various accretion as well as shock parameters.\\
However, other than the Paczy\'nski-Wiita potential,
a number of modified Newtonian potentials of various forms
are also available in the literature
which accurately approximate some general relativistic features of rotating 
accretion around  Schwarzschild black holes.
Such potentials may be called `pseudo - Schwarzschild' potentials because they 
mimic the space-time around a non-rotating/slowly
rotating compact object.
Recently Das and  Sarkar$^6$ (DS hereafter) has shown that 
though these so called
pseudo potentials were originally proposed to mimic the relativistic 
effects manifested in disc accretion, it is quite reasonable to use most 
of these potentials in studying various dynamical and thermodynamic 
quantities also for spherical accretion on to 
Schwarzschild black holes and it was established that along with the 
Paczy\'nski and Wiita potential,$^5$ two other potentials proposed by 
Artemova et. al.$^1$ also provide reasonably good approximation to the
complete general relativistic description of transonic, spherically 
symmetric accretion on to a Schwarzschild black hole.\\
\noindent
Remembering  that the free-fall acceleration plays a very crucial
role in Newtonian gravity, Artemova et. al.$^1$  proposed the following
two 
potentials to study disc accretion around a non-rotating black hole.
The first potential proposed by them produces exactly the
same value of the free-fall
acceleration of a test particle at a given value of $r$ as is obtained
for a test particle at rest with respect to the Schwarzschild reference
frame, and is given by
$$
\Phi_{1}=-1+{\left(1-\frac{1}{r}\right)}^{\frac{1}{2}}
\eqno{(1a)}
$$
The second one gives the value of the free fall acceleration that is equal
to the value of the covariant component of the three dimensional free-fall
acceleration vector of a test particle that is at rest in the Schwarzschild
reference frame and is given by
$$
\Phi_{2}=\frac{1}{2}ln{\left(1-\frac{1}{r}\right)}
\eqno{(1b)}
$$
For Eq. 1(a-b), $r$ represents the radial co-ordinate scaled in the unit 
of $r_g$.
Efficiencies produced by $\Phi_1$ and $\Phi_2$ are $-0.081$ and $-0.078$
respectively. As both $\Phi_1$ and
$\Phi_2$ stems from the
consideration of free fall acceleration and calculates the dependence of
free fall acceleration on radial distance in the Schwarzschild metric
(which
describes a spherically symmetric gravitational field in vacuum), it
appears to be
quite justified to use those potentials to study spherically symmetric
accretion.\\
\noindent
Owing to the fact that $\Phi_1$ and $\Phi_2$ may be used to mimic the 
spacetime around a spherically accreting non-rotating black hole quite nicely,
we believe that along with the calculation presented in D99, it is equally 
important to investigate the various features of the accretion powered 
spherical outflow using these two potentials. In this paper we precisely do this,
first we formulate and solve the equations governing the inflow and outflow 
using  $\Phi_1$ and $\Phi_2$ (in presence of a steady, standing spherical 
shock as described in previous paragraphs) and then self-consistently 
connect the accretion and wind solutions to calculate what fraction of the 
accreting material is being blown as wind. Also we study the dependence of
this fraction on various flow parameters. The plan of this paper is as follows. In next section we describe how to formulate
and solve the governing equations. In \S 3, we present our results. In \S 4, the
results will be reviewed and conclusions will be drawn.\\

\section{Governing equations and solution procedure}

\subsection {Inflow model}

Hereafter we will refer to these two potentials as $\Phi_i$ in general
where $\{i=1,2\}$ would correspond to $\Phi_1$ (Eq. (1a)) and $\Phi_2$
(Eq. (1b)) respectively. We assume that a Schwarzschild type black hole
spherically accretes fluid
obeying  a polytropic equation of state. The density of the fluid is $\rho(r)$,
$r$ being the radial distance measured in the
unit of Schwarzschild radius $r_g$. We also assume that the accretion rate
(scaled in the unit of
Eddington rate ${\dot M}_{Edd}$) is not a function
of $r$ and we ignore the
self-gravity of the flow.
For simplicity of calculation, we choose the geometric unit
where the unit of length is scaled in units of $r_g$, units of velocity
in units of $c$ . All other physically relevant
quantities can be expressed likewise. We also set $ G=c=1$ in the system of
units used here.
It is to be mentioned here that one fundamental criterion
for formation of hydrodynamic outflows is that
the outflowing wind should have a positive Bernoulli constant which means
that the matter in the post-shock region is able to escape
to infinity. However, positiveness in Bernoulli's constant may lead to another
situation as well where shock may quasi-periodically originate
at some certain radius and propagate outwards without formation of
outflows. So formation of outflows is one of the possible scenarios
when we focus on the positive energy solutions.
In this paper we concentrate
only on solutions producing outflows. 
Another assumption made in this paper
is to treat the accreting as well as post-shock matter as a single temperature
fluid, the temperature of which is basically characterized by proton 
temperature. \\
\noindent
For any pseudo-Schwarzschild potential $\Phi_i$, the
equation of motion for spherically accreting  material
onto the accretor
is given by 
$$
\frac{{\partial{u}}}{{\partial{t}}}+u\frac{{\partial{u}}}{{\partial{r}}}+\frac{1}{\rho}
\frac{{\partial}P}{{\partial}r}+{\Phi_i}^{'}=0
\eqno{(2a)}
$$
where symbols have their
usual meaning.
The first term of Eq. 2(a) is the Eulerian time derivative of the
dynamical velocity at a given $r$, the second term 
is the `advective' term, the third term 
is the
momentum deposition due to pressure gradient and the
final term is due to the gravitational acceleration
for a particular $i$th potential $\Phi_i$. The continuity
equation can be written as 
$$
\frac{{\partial}{\rho}}{{\partial}t}+\frac{1}{r^2}\frac{{\partial}}{{\partial}r}\left({\rho}ur^2\right)=0
\eqno{(2b)}
$$
For a polytropic equation of state, i.e., $p=K{\rho}^{\gamma}$, 
the steady state
solutions 
of eqn. 2(a) and eqn. 2(b) are\\
1) Conservation of specific energy ${\cal E}$ of the flow:
$$
{\cal E}=\frac{u^2}{2}+\frac{a^2}{{\gamma_{in}}-1}+\Phi_i
\eqno{(3a)}
$$
where $\gamma_{in}$ is the adiabatic index of the inflow (accretion)
and\\ 
\noindent
2) Conservation of Baryon number (or accretion rate ${\dot M}_{in}$):
$$
{\dot M}_{in}=4{\pi}{\rho}ur^2
\eqno{(3b)}
$$
Using ${\dot {\cal M}}_{in}$  as the entropy accretion rate where 
${\dot {\cal M}}=
{\dot M}_{in}{\gamma_{in}}^{\frac{1}{\gamma_{in}-1}}K^{\frac{1}
{\gamma_{in}-1}}$, 
eqn. 3(b) can be rewritten as (see DS and references therein):
$$
{\dot {\cal M}}_{in}=4{\pi}a^{\frac{2}{\gamma_{in}-1}}ur^2
\eqno{(3c)}
$$
It is now quite straightforward to derive the spatial 
gradient of dynamical velocity $\left(\frac{du}{dr}\right)_i$ 
for flow in any particular $i$th potential $\Phi_i$ as
$$
{\left(\frac{du}{dr}\right)}_i=\frac{\frac{2a^2}{r}-
{\Phi_i}^{'}}{u-\frac{a^2}{u}}
\eqno{(4a)}
$$
where $\left|{{{{{\Phi}^{'}}_{i}}}}\right|$
denotes the absolute value of the
space derivative of $\Phi_i$, i.e.,
$$
\left|{{{{{\Phi}^{'}}_{i}}}}\right|=\left|{\frac{d{\Phi_i}}{dr}}\right|
$$
Since the flow is assumed to be smooth everywhere, if
the denominator of Eq. 4(a)  vanishes at any radial distance
$r$, the numerator must also vanish there to maintain the
continuity of the flow. One therefore arrives at the so
called `sonic point 
conditions'  by simultaneously making
numerator and denominator of Eq. 4(a) equal to zero and
the sonic point conditions can be expressed as follows
$$
{u^i}_c={a^i}_c=\sqrt{\frac{r^i_c}{2}{{\Phi_i}^{'}}{{\Bigg{\vert}}}_c}
\eqno{(4b)}
$$
where superscript $i$ denotes the specific
value of sonic quantities for a particular $i$th
potential $\Phi_i$, and ${\Phi^{'}_i}{\bigg{\vert}}_c$ is the 
value of
$\left(\frac{d{\Phi_i}}{dr}\right)$ evaluated at the corresponding sonic point
$r^i_c$. The value of sonic point $r^i_c$ for any $i$th potential 
$\Phi_i$ can be obtained
by algebraically solving the following equation
$$
{\cal E}-\frac{1}{2}\left(\frac{\gamma_{in}+1}{\gamma_{in}-1}\right)r^i_c
{{\Phi_i}^{'}}{{\Bigg{\vert}}_c}
-{\Phi_i}{\Bigg{\vert}}_c=0
\eqno{(4c)}
$$
where ${\Phi_i}{\bigg{\vert}}_c$ is the value of $i$th potential at the
corresponding sonic point $r^i_c$. 
 Similarly, the value of $\left(\frac{du}{dr}\right)_i$ for any $\Phi_i$
at its
corresponding sonic point $r^i_c$ can be obtained by
solving the following quadratic equation:
$$
\left(1+\gamma_{in}\right){{\left(\frac{du}{dr}\right)}^2}_{c_,i}+
2.829\left(\gamma_{in}-1\right)\sqrt\frac{{{\Phi_i}^{'}}
{{\Bigg{\vert}}_c}}{r^{i}_c}{\left(\frac{du}{dr}\right)}_{c_,i}
$$
$$
+\left(2{\gamma_{in}}-1\right)
\frac{
{{\Phi_i}^{'}{\Bigg{\vert}}_c}}
%{{r^i_c}}
{{r^i_c}}
+{{\Phi_i}^{''}}{{\Bigg{\vert}}_c}=0
\eqno{(4d)}
$$
where ${{\Phi_i}^{''}}{{\Bigg{\vert}}_c}$
is the value of $\frac{d^2\Phi_i}{dr^2}$  
at the corresponding
critical point $r^i_c$.\\
\noindent
One can simultaneously solve eqn. 3(a) and eqn. 3(b)
(alternatively, eqn. 3(a) and eqn. 3(c))
for any specific $\Phi_i$ for a
fixed value of ${\cal E}$ and $\gamma_{in}$
to obtain various dynamical and thermodynamic flow quantities. 
In this work we normally use the value of $\gamma_{in}$
to be $\frac{4}{3}$. Though far away from
the black hole, optically thin accreting plasma
may not be treated as
ultra-relativistic fluid
(by the term `ultra-relativistic' and `purely non-relativistic'
we mean a flow with
$\gamma=\frac{4}{3}$ and $\gamma=\frac{5}{3}$ respectively,
according to the terminology used in
Frank et. al.$^7$) in general, close to the
hole it always advects with enormously large radial velocity and
could be well-approximated as ultra-relativistic flow.
As because our main region of interest,
the shock formation zone, normally  lies close to the black hole
(a few tens of $r_g$ away from the hole
or sometimes even less, see results and figures in \S 3),
we believe that
it is fairly justifiable to assign the value $\frac{4}{3}$ for
$\gamma_{in}$ in our work.
However, to rigorously model a real flow without any assumption,
a variable polytropic index
having proper functional dependence on radial distance
(i.e., $\gamma_{in}~\equiv~\gamma_{in}(r)$) may be considered
instead of using a constant $\gamma_{in}$,
and equations of
motion may be formulated accordingly, which we did not attempt
here for the sake of simplicity.
Nevertheless, we keep our option open for values of $\gamma_{in}$
other than $\frac{4}{3}$ as well and investigated the outflow
for an range of values of $\gamma_{in}$ for a specific
value of ${\cal E}$ and ${\dot M}_{Edd}$ (Fig. 4, \S 3.3).
The same kind of investigations could be performed for a variety of values of
${\cal E}$ and ${\dot M}_{Edd}$ and a set of results may be obtained
with  a wide range of values of $\gamma_{in}$ which tells that our
calculation is not restricted to the value $\gamma_{in}~=~\frac{4}{3}$
only; rather the model is general enough to deal with all possible
value of $\gamma_{in}$ for polytropic accretion.

\subsection{Shock parameters and the outflow model}

As already mentioned, a steady, collision-less shock forms
due to the instabilities in the plasma flow (see D99 and references therein
for details of 
shock formation mechanism). We take this shock surface to be the effective 
physical barrier around the black hole which might be responsible to 
generate the outflow.
We assume that the presence of a collision-less
steady standing spherical shock discussed in this work
may randomize
the directed in-fall motion
and at the shock surface the individual components of the total energy
of the flow (which is a combination of the kinetic, thermal and
gravitational energy) gets rearranged in such a manner that the thermal energy
of the post-shock matter dominates (due to enormous shock generated 
post-shock proton
temperature) over the gravitational attraction of the accretor
and a part of the in-falling material is driven by thermal pressure
to escape to infinity as wind. In \S 3. we show that for any
shock solution, 
the mass-loss rate normally co-relates with
the post-shock proton temperature which essentially supports the
validity of our assumption.
We also assume 
the effective thickness of the shock $\Delta_{sh}$ to be 
small enough compared
to the shock standoff distance $r_{sh}$,
and that the relativistic particles encounter a full shock compression
ratio while crossing the shock.\\
\noindent
At the shock, density of matter will shoot up and inflow velocity will drop
abruptly. If $({\rho}_{-}, u_{-})$ and $({\rho}_{+}, u_{+})$ are the pre-
and post-shock densities and velocities respectively at the shock surface,
then
$$
\frac{{\rho}_{+}}{{\rho}_{-}} = R_{comp} = \frac{u_{-}}{u_{+}}
\eqno {(5)}
$$
where $R_{comp}$ is the shock compression ratio.
For high shock Mach number solution,
the expression for $R_{comp}$
can be well approximated as
$$
R_{comp} = 1.44{M_{sh}}^{\frac{3}{4}}\\
\eqno{(6)}
$$
where $M_{sh}$ is the shock Mach number and Eq. (5) holds for
$M_{sh} \gsim 4.0$.$^8$ \\
In terms of various accretion parameters,
shock location can be computed as (D99):
$$
r_{sh} = \frac{3{\sigma_{pp}}{\dot M}_{Edd}}{4{\pi}{u_{sh}}^2}\left(
\frac {1~-~2.4{M_{sh}}^{-0.68}}{1~-~3.2{{M_{sh}}^{-0.62}}}\right)
\eqno{(7)}
$$
where ${\sigma}_{pp}$ is the collision cross section for relativistic
protons, $u_{sh}$ and $M_{sh}$ are the dynamical flow velocity and the Mach
number attained at the shock location, ${\dot M}_{Edd}$ is the mass
accretion rate scaled in units of Eddington rate. One can understand that \\
$$
\left(u_{sh}, M_{sh}\right)\equiv{\bf \zeta}\left({\cal E},{\dot M}_{Edd},
\gamma_{in}\right)
\eqno{(7a)}
$$
where ${\bf {\zeta}}$ has some complicated non-linear functional form
which cannot be evaluated analytically, but the value of $u_{sh}$ and
$M_{sh}$ can easily be obtained in terms of $\left\{{\cal E},{\dot M}_{Edd},
\gamma_{in}\right\}$ by numerically solving Eq. 3(a-b) and Eq. 7 (with the
help of Eq. 4(a-d)) simultaneously. Hence one can write \\
$$
r_{sh}\equiv{\bf {\xi}}\left({\cal E},{\dot M}_{Edd},\gamma_{in}\right)
\eqno{(7b)}
$$
where $\xi$ has some functional form other than that of $\zeta$.\\
\noindent
In ordinary stellar mass-loss computations (Tarafder,$^9$ and references 
therein),
the outflow is assumed
to be isothermal till the sonic point. This assumption is probably justified,
since
copious photons from the stellar atmosphere deposit momenta on the slowly 
outgoing
and expanding outflow and possibly make the flow close to isothermal. This
 need
not be the case for outflow from black hole candidates. Our effective 
boundary layer,
being close to the black hole, are very hot
and most of the photons emitted
may be swallowed by the black hole itself instead of coming out of the region
and
depositing momentum onto the outflow. Thus, the outflow could be cooler than
the isothermal flow in our case. We choose polytropic outflow with a different
polytropic index ${{\gamma}_o} < {\gamma_{in}}$ due to momentum deposition.
In our calculation we also assume that essentially the post-shock
fluid pressure and the post-shock proton temperature controls the
wind formation as well as the barionic matter content of the outflow.\\
\noindent
The adiabatic post-shock sound speed $a_{sh}^{+}$ and the
post-shock temperature $T_{sh}^{+}$ (which is basically the
temperature of the protons according to our one-temperature fluid
approximation) can be calculated as:
$$
a_{sh}^{+}=\sqrt{\frac{{\gamma_o}p_{sh}^{+}}{\rho_{sh}^{+}}}
\eqno{(8a)}
$$
and
$$
T_{sh}^{+}=\frac{{\mu}{m_p}p_{sh}^{+}}{{\kappa}{\rho_{sh}^{+}}}
\eqno{(8b)}
$$
where $p_{sh}^{+}$ and $\rho_{sh}^{+}$ are the post-shock pressure and density
of the flow at shock location $r_{sh}$ respectively.
For low energy accretion (`cold' inflow, so to say) which is appropriate to 
produce
a high shock Mach number solution, one can assume that the pre-shock thermal 
pressure
($p_{sh}^{-}$) may be neglected compared to its post-shock value
($p_{sh}^{+}$) and to the pre-shock ram pressure
$\left({\rho_{sh}^{-}}\left(u_{sh}^{-}\right)^2\right)$.
One can obtain the value
of $p_{sh}^{+}$
using Eq. (5-6) and from the total pressure
balance condition at shock as,
$$
p_{sh}^{+} = \left({u_{sh}^{+}}\right)^2r_{sh}\left({\frac{R_{comp} - 1}{R_{comp
}}}\right)
\eqno{(9)}
$$
Combining Eq. (5-9), post-shock sound velocity and temperature
obtained at the shock surface can be rewritten as:
$$
a^+_{sh}=3.54\left(r_{sh}u^+_{sh}\right)^{1.5}
\sqrt{{\frac{\gamma_o}{{\dot M}_{Edd}}}
\left(
\frac{1.44M^{\frac{3}{4}}_{sh}-1}{M^{\frac{3}{4}}_{sh}}\right)}
\eqno{(10a)}
$$
and \\
$$
T^+_{sh}=\frac{4{\pi}{\mu}m_p}{{\kappa}{{\dot M}_{Edd}}}
\left(r_{sh}u^+_{sh}\right)^{3}
\left(
\frac{1.44M^{\frac{3}{4}}_{sh}-1}{M^{\frac{3}{4}}_{sh}}\right)
\eqno{(10b)}
$$
General form of the conservation equations 
governing the polytropic outflow will be the 
same as Eq. (3a - 3c) with different polytropic index and total specific 
energy and entropy (${\cal E}~>~{\cal E}^{'}$,
${\dot {\cal M}}_{in}~<~{\dot {\cal M}}^{'}$ and $\gamma_{in}~>~\gamma_o$,
see D99). So we can write:
$$
{\cal E}^{'}=\frac{\left({u^o}\right)^2}{2}+\frac{\left({a^o}\right)^2}
{{\gamma_{o}}-1}+\Phi_i
\eqno{(11a)}
$$
$$
{\dot M}_{out}=4{\pi}{\rho^o}u^or^2
\eqno{(11b)}
$$
and\\
$$
{\dot {\cal M}}^{'}=4{\pi}\left(a^o\right)^{\frac{2}{\gamma_o-1}}u^or^2
\eqno{(11c)}
$$
where where ${\cal E}^{'}$
is the specific energy 
of the outflow which is also assumed to be
constant throughout the flow and
${\dot {\cal M}}^{'}={\dot M}_{out}{\gamma_{o}}^{\frac{1}{\gamma_{o}-1}}
{K^{o}}^{\frac{1}{\gamma_{o}-1}}$ is the entropy accretion rate of the outflow.
$\gamma_o~<~\gamma_{in}$ as already mentioned. Any sub/ super-script indicates 
that the quantities are measured for the outflow. Like Eqs. (4a - 4d), one can
easily write the sonic point conditions and the velocity gradient of the
outflow as:
$$
{\left(\frac{du^o}{dr}\right)}_i=\frac{\frac{2\left(a^o\right)^2}{r}-
{\Phi_i}^{'}}{u^o-\frac{\left(a^o\right)^2}{u^o}}
\eqno{(12a)}
$$
$$
\left({u^i}_c\right)^o=\left({a^i}_c\right)^o=
\sqrt{\frac{\left(r^i_c\right)^o}{2}{{\Phi_i}^{'}}{{\Bigg{\vert}}}_c}
\eqno{(12b)}
$$
$$
{\cal E}^{'}-\frac{1}{2}\left(\frac{\gamma_{o}+1}{\gamma_{o}-1}\right)
\left(r^i_c\right)^o
{{\Phi_i}^{'}}{{\Bigg{\vert}}_c}
-{\Phi_i}{\Bigg{\vert}}_c=0
\eqno{(12c)}
$$
and\\
$$
\left(1+\gamma_{o}\right){{\left(\frac{du^o}{dr}\right)}^2}_{c_,i}+
2.829\left(\gamma_{o}-1\right)\sqrt\frac{{{\Phi_i}^{'}}
{{\Bigg{\vert}}_c}}{\left(r^{i}_c\right)^o}{\left(\frac{du^o}{dr}\right)}_{c_,i}
$$
$$
+\left(2{\gamma_{o}}-1\right)
\frac{
{{\Phi_i}^{'}{\Bigg{\vert}}_c}}
%{{\left(r^i_c\righth)^o}}
{{\left(r^i_c\right)}}
+{{\Phi_i}^{''}}{{\Bigg{\vert}}_c}=0
\eqno{(12d)}
$$
We then define the mass outflow rate $R_{\dot m}$ as:
$$
R_{\dot m}=\frac{{\dot M}_{out}}{{\dot M}_{in}}
\eqno{(13)}
$$
It is obvious from the above discussion that $R_{\dot m}$ should have some
complicated non-linear functional dependence on the following accretion and
shock parameters:
$$
{R}_{\dot m} \equiv  {\bf {\Psi}}\left({\cal E},{{\dot M}_{Edd}},{r_{sh}},M_{sh},
{R_{comp}},{\gamma_{in}},{{\gamma}_o}\right)
\eqno{(14a)}
$$
As $r_{sh}, M_{sh}$ and $R_{comp}$ can be found in terms of ${\cal E},
{\dot M}_{Edd}$ and $\gamma_{in}$ only, ultimately it turns out that:
$$
{R}_{\dot m} \equiv {\bf \Omega}\left({\cal E}, {\dot M}_{Edd}, \gamma_{in}, \gamma_o
\right)
\eqno{(14b)}
$$
Where ${\bf {\Omega}}$ has some complicated functional form which cannot be
evaluated analytically.

\subsection{Simultaneous solution of inflow-outflow equations}

\noindent
In this work, we are interested in finding
the ratio of ${\dot M}_{out}$ to ${\dot M}_{in}$ (Eq. (13)), and
not the  explicit value of ${\dot M}_{out}$.
Also note that the primary goal of our present work was to compute the
outflow rate and to investigate its dependence on various inflow parameters but
not to study the collimation procedure of the outflow.\\
Before we proceed in detail, a general understanding of the transonic
inflow outflow system in the present case is essential to understand the basic scheme
 of the
solution procedure. Let us consider the transonic accretion first. Infalling matter
becomes supersonic after crossing a saddle-type sonic point, the location of which is
determined by $\left\{{\cal E},~{\dot M}_{Edd},~\gamma_{in}\right\}$. 
This supersonic
flow then
encounters a shock (if present), location of which ($r_{sh}$) is determined from
Eq. (7).
At the shock surface, part of the incoming matter, having a higher entropy density
(because shock in a fluid flow generates entropy),
is likely
to return as wind through a sonic point other than the point
through which it just entered.
Thus a combination of transonic topologies,
one for the inflow and other for the outflow
(passing through a {\it different}
 sonic point and following topology completely
different that of the `self-wind' of the accretion), is required to obtain a full
solution. So it turns out that finding a complete set of self-consistent
inflow outflow solutions ultimately boils down to locate
the sonic point of the polytropic outflow and the mass
flux through it. Thus a supply of parameters ${\cal E}, {{\dot M}_{Edd}}$,
${\gamma_{in}}$ and $\gamma_o$
make a self-consistent computation of ${R_{\dot m}}$ possible.
Here $\gamma_o$ is supplied as free parameter because the self-consistent
computation
of $\gamma_o$ directly using ${\cal E}, {{\dot M}_{Edd}}$ and
$\gamma_{in}$ has not been attempted in this work;
instead we put a constrain that $\gamma_o < \gamma_{in}$ always and for any value of
$\gamma_{in}$.
In reality, $\gamma_o$ is directly related to the heating and cooling 
processes taking place 
in the outflow.\\
\noindent
We obtain the inflow sonic point $r_c$ by solving Eq. (4c).
Using the fourth order
the Runge Kutta method, $u(r)$, $a(r)$ and
the inflow Mach number
$\left[\frac{u(r)}{a(r)}\right]$ are computed along the  inflow from the
{\it inflow} sonic point $r_c$ till
the position where the shock forms. The shock location is calculated
by solving Eq. (7).
Various shock parameters
(i.e., density, pressure etc at the shock surface) are then computed
self-consistently.\\
For outflow, with the known value of ${\cal E}^{\prime}$
and $\gamma_o$, it is easy to compute the location of the outflow sonic point 
$r_c^o$ 
from Eq. (12c). At the
outflow sonic point, the outflow velocity $u^o_c$ and polytropic sound 
velocity $a_c^o$ is computed
from Eq. (12b). Using Eq. (12a) and (12d),
$\left(\frac{du^o}{dr}\right)$
and $\left(\frac{du^o}{dr}\right)_c$ is computed as was
done for the inflow. Runge
-Kutta method is then
employed to integrate from the {\it outflow} sonic point $r_c^o$ towards the black hole
to
find out the outflow velocity $u^o$ and density $\rho^o$ at the shock location.
The mass outflow rate $R_{\dot M}$ is then computed using Eq.(13).

\section{Results}

\subsection{Shock location as a function of ${\cal E}$ and
${\dot M}_{Edd}$ and related post-shock quantities}

\noindent
For a particular value of ${\cal E}$,
${\dot M}_{Edd}$ and $\gamma_{in}$,
the shock location (measured from the black hole in units of $r_g$) can be
calculated using Eq. (7). As $u_{sh}$ and $M_{sh}$ is a function of {${\cal E}$
,
${\dot M}_{Edd}$, and $\gamma_{in}$}, $r_{sh}$ will also change with the change
of
any of these accretion parameters. 
In figure 1, we show the variation of $r_{sh}$ as a function of 
${\cal E}$ for three different values of ${\dot M}_{Edd}$. While ${\cal E}$ is
plotted along the X axis, shock location (in logarithmic scale) is plotted 
along the Y axis for a fixed value of $\gamma_{in}(=\frac{4}{3})$. Three
different curves drawn by solid lines for $\Phi_2$ and dotted lines for
$\Phi_1$, are plotted for three different values of ${\dot M}_{Edd}$
(=0.25, 1.0, 1.75). The lowermost curves (for each of the pseudo-potentials)
corresponds to the value ${\dot M}_{Edd}=0.25$ as shown in the figure.
For both $\Phi_1$ and $\Phi_2$, other two curves from bottom to top,
correspond to ${\dot M}_{Edd}$ equal to $1.0$ and $1.75$ respectively.
For any pseudo-potential one should note that different curves terminate (in the
direction of increasing ${\cal E}$) at different points which indicates that
shock formation is not a generic phenomena, i.e., shock does not
form for any value of ${\cal E}$, ${\dot M}_{Edd}$, and $\gamma_{in}$,
rather a specific region of parameter space spanned by $\left\{
{\cal E}, {\dot M}_{Edd}, \gamma_{in}\right\}$ allows shock formation. 
Both sub- as well as super-Eddington accretion allows shock formation 
as shown in the figure. For any pseudo-potential, while the shock location
non-linearly anticorrelates with ${\cal E}$ (for a fixed value of 
${\dot M}_{Edd}$ and $\gamma_{in}$), it correlates (non-linearly) with
${\dot M}_{Edd}$ (for a fixed value of ${\cal E}$ and $\gamma_{in}$).
The maximum value of ${\cal E}$ for which shock may form for a fixed
value of ${\dot M}_{Edd}$ and $\gamma_{in}$, increases with increase of
the accretion rate of the flow. It is observed that (figuratively not shown
in the paper) the shock location also non-linearly anti-correlates with
$\gamma_{in}$ which means that for both $\Phi_1$ and $\Phi_2$, 
non-relativistic 
super-Eddington
accretion with low specific energy of the flow is a proper combination
to produce the shock closest to the black hole (this result has significant 
importance in studying the amount of barionic load in the wind, see 
\S 3.3). Once the shock location is known, one can easily calculate 
any post-shock or shock related (the shock compression ratio for 
example) quantity using the corresponding equations derived in \S 2.2.
One important physical quantity of our interest is the post-shock 
proton temperature of the flow (which is practically the temperature of the 
outflow according to our one-fluid approximation, see \S 2.1) $T^{+}_{sh}$ which can be
computed using Eq. 10(b). We have seen that for any pseudo-potential,
$T^{+}_{sh}$ correlates with ${\cal E}$ as well as with $\gamma_{in}$,
hence high energy pure-nonrelativistic flow would produce hotter outflow
as well as it is evident that the closer the shock forms to the 
accretor, the higher becomes the post-shock proton temperature of
the flow. \\
\noindent 
Whatever observations are presented in the above paragraph, is commonly 
applicable for both $\Phi_1$ and $\Phi_2$. However, there are a number of
differences in magnitude of post-shock quantities (even in the shock location)
observed when the flow is studied for two different pseudo-potentials. 
From figure 1, one can observe that for the same values of ${\cal E}$,
${\dot M}_{Edd}$ and $\gamma_{in}$, the shock forms relatively 
{\it closer} to the black hole for $\Phi_2$ compared to the case for 
$\Phi_1$. If $r_{sh}{\Bigg{\vert}}_{\Phi_i}$ is the shock location
obtained using any particular $i$th pseudo-potential, we see that:
$$
r_{sh}{\Bigg{\vert}}_{\Phi_2}~>~r_{sh}{\Bigg{\vert}}_{\Phi_1}
\eqno{(15)}
$$
for a fixed value of $\left\{{\cal E}, {\dot M}_{Edd}, \gamma_{in}\right\}$.
This deviation is more prominent for flows with higher specific energy
and the following quantity
$$
{\delta}r_{sh}=\left[r_{sh}{\Bigg{\vert}}_{\Phi_2}
-r_{sh}{\Bigg{\vert}}_{\Phi_1}\right]
$$
decreases in the direction of low ${\cal E}$. Also we note that the 
maximum value of ${\cal E}$ for which the shock forms for a fixed
value of ${\dot M}_{Edd}$ and $\gamma_{in}$ (let us define that 
energy as ${\cal E}_{max}{\Bigg{\vert}}_{\Phi_i}$ for any $i$th 
pseudo-potential), is higher for flows in $\Phi_2$, i.e., 
$$
{\cal E}_{max}{\Bigg{\vert}}_{\Phi_2} ~>~{\cal E}_{max}{\Bigg{\vert}}_{\Phi_a}
$$
but unlike ${\delta}r_{sh}$, ${\delta}{\cal E}_{max}$ 
does not show any specific dependence on inflow parameters.   
It is also observed that (not shown in the figure) for a fixed values
of energy and accretion rate, if we study the shock location as a 
function of the polytropic index of the inflow, the inequality presented
in equation (15) is still maintained and ${\delta}r_{sh}$ decreases as the flow
tends to its pure non-relativistic limit. As the post-shock proton 
temperature $T^{+}_{sh}$ is inversely proportional with the shock
location, we find that,
$$
T^{+}_{sh}{\Bigg{\vert}}_{\Phi_2}~<~T^{+}_{sh}{\Bigg{\vert}}_{\Phi_1}
\eqno{(16)}
$$
for both the cases when:\\
\noindent
a) ${\cal E}$ is being varied keeping ${\dot M}_{Edd}$ and $\gamma_{in}$ 
constant.\\
\noindent
b) Flow is being studied as a function of $\gamma_{in}$ for fixed 
values of ${\cal E}$ and ${\dot M}_{Edd}$.\\
\noindent
However, ${\delta}T^{+}_{sh}$ decreases with decrease in ${\cal E}$
but with increase in  $\gamma_{in}$.

\subsection{Combined integral curves of motion}

\noindent
Figure 2a and 2b show  two typical solutions which combines the accretion and
outflow for flows in $\Phi_1$ and $\Phi_2$ respectively.
The accretion parameters used are ${\cal E}$ = 0.0003, ${\dot M}_{Edd}$
=0.5 
and $\gamma_{in} = \frac{4}{3}$
corresponding to ultra-relativistic inflow.
For both of the figures 2a and 2b,
the solid curve AB represents the pre-shock region of the
inflow and the solid vertical line BC with double arrow at $r_{sh}$
represents
the shock transition. Shock locations
(9.9 $r_g$ for the Fig. 2a and 10.6 $r_g$ for the Fig. 2b) 
is obtained using the Eq.(7)
for a particular set of inflow parameters mentioned above.
Three dotted curves show the three different outflow branches corresponding 
to the three different adiabatic indices $\gamma_o$ of the outflow. From left 
to right, the values of $\gamma_o$ are 1.3, 1.25 and 1.2 respectively 
with respective mass-outflow rates as  
$1.558\times10^{-3},~7.1487\times10^{-5}$ and $6.249\times10^{-7}$
(for Fig. 2a) and 
$1.588\times10^{-3},~7.47\times10^{-5}$ and $6.765\times10^{-7}$ 
(for Fig. 2b)
 respectively which indicates 
that for a given value of ${\cal E}$, ${\dot M}_{Edd}$ and 
$\gamma_{in}$, $R_{in}$ correlates with $\gamma_o$.
It is evident from the figure that the outflow moves along the 
solution curves in a completely different way to that of the `self-wind' 
solution of the inflow (solid line marked by CD in Fig. 2a and 2b).
Also, the sonic points for 
all the outflowing branches are different to those of the accretion `self-wind' 
system which is designated as $P_s$. While $P_s=1250.833~r_g$ for 
for Fig. 2a and $1251.665~r_g$ for Fig. 2b, the sonic points 
of the outflowing branches corresponding to 
$\gamma_o=1.3,1.25$ and $1.2$
are 
$1528.55~r_g,~2084.033~r_g,~2917.31~r_g$ for Fig. 2a 
and $1529.32~r_g,~2084.733~r_g,~2917.95~r_g$ for Fig. 2b
respectively, which indicates that the 
outflow sonic point {\it increases} with a {\it decrease} in the adiabatic
index of the {\it outflow} and thus the wind starts with a very low bulk velocity which is 
why the mass-loss rate decreases. It is also observed that the sonic point of the 
accretion-`self-wind' system is, in general, located {\it closer} to the event
horizon compared to the outflow sonic point for {\it all values} of ${\cal E},
{\dot M}_{Edd}, \gamma_{in}$ and $\gamma_o$.\\
\noindent
Combining the informations obtained from the Fig. 2a and 2b, one can observe
that for a fixed set of values of $\left\{{\cal E}, {\dot M}_{Edd},
\gamma_{in}, \gamma_o\right\}$, the inflow sonic point
(shown by $P_s$ in the figures) and the outflow sonic points
(for various $\gamma_o$) for flows in $\Phi_2$, is greater than 
those quantities obtained for flow in $\Phi_1$. Whereas the shock
Mach number $M_{sh}$ for $\Phi_1$ has a higher value compared to 
$M_{sh}$ for $\Phi_2$. 
This is obvious because the potential $\Phi_2$ is steeper 
compared to $\Phi_1$ (see DS) hence $\Phi_2$ produces relatively higher 
acceleration and the accreting material gains more kinetic energy 
relatively faster to become supersonic at a distance relatively greater 
compared to the flow in $\Phi_1$ and 
as the sonic
point of the inflow resides relatively further away from the 
event horizon, accreting material becomes supersonic at an earlier stage
and the rate of increment of the ratio of dynamical (mechanical) to the 
thermal energy content of the accretion increases; also another contributing
factor is the location of the shock, Eq. (15) also tells that $M^{+}_{sh}$ 
for $\Phi_2$, in reality, should be less compared to its value for
$\Phi_1$. However, for any set of $\left\{{\cal E}, {\dot M}_{Edd},
\gamma_{in}, \gamma_o\right\}$, the mass outflow rate $R_{\dot m}$ for
$\Phi_1$ is less than that obtained in $\Phi_2$. We will come
to this point in detail in next sub-section.

\subsection{Dependence of $R_{\dot m}$ on the specific energy of the flow}

\noindent
In figure 3a, we have plotted the variation of $R_{\dot m}$ with the 
specific energy of the flow ${\cal E}$ for a fixed values of
$\gamma_{in}~(=\frac{4}{3})$ and $\gamma_o~(=1.3)$ and for a number of
values (shown in the figure)
of the accretion rate (scaled in units of Eddington rate and
shown in the figure as ${\dot M}_{Edd}$) of the inflowing material
for flows in $\Phi_1$ (dotted lines), as well as for flows in $\Phi_2$
(solid lines). Any dotted line accompanying a solid line marked
by a specific value of ${\dot M}_{Edd}$ corresponds to the same
accretion rate as that of the solid lines. 
We observe that for a fixed 
value of ${\dot M}_{Edd}$, $\gamma_{in}$ and $\gamma_o$, the mass-outflow 
rate non-linearly correlates with ${\cal E}$ reason for which might
be as follows:\\
\noindent
As ${\cal E}$ increases, $r_{sh}$ decreases
(see Fig. 1, \S 3.1) and the post-shock bulk velocity of the flow 
$u_{sh}^{+}$ as well as the post-shock density $\rho_{sh}^{+}$
increases. 
The outflow rate, which is the product of three quantities $r_{sh}$,
$\rho_{sh}^{+}$ and $u_{sh}^{+}$ (see Eq. (11b)),
increases in general due to the combined `tug of war'
of these three quantities.
Moreover, the closer the shock is to the black hole, the greater the
amount
of gravitational potential will be available to be put onto the relativistic
protons to provide stronger outward pressure  and the closer the shock forms
to the black hole, the higher is the post-shock proton temperature (the
effective characteristic outflow temperature) and the higher is the
amount of outflow (as the wind is observed to be strongly 
thermally driven, see discussion below).
Thus the mass-outflow rate increases with ${\cal E}$ because for a particular 
set of fixed values of ${\dot M}_{Edd}$, $R_{\dot m}$ is proportional to 
${\dot M}_{out}$ which increases with ${\cal E}$. \\
\noindent
The unequal gaps between the curves marked with different ${\dot M}_{Edd}$ 
in the figure imply that when the inflow specific energy is kept 
constant, $R_{\dot m}$ non-linearly increases with the accretion rate 
of the infalling material. This is because, as ${\cal E}$ is kept constant
while ${\dot M}_{Edd}$ is varied, the amount of infalling energy 
converted to produce the high-energy protons is also fixed. So the higher
the value of ${\dot M}_{Edd}$, the larger the distance of the shock surface 
from the event horizon and the outflowing matter feels a low inward 
gravitational pull, the result of which is the non-linear correlation
of $R_{\dot m}$ with  ${\dot M}_{Edd}$. In other words, for a fixed 
value of $\gamma_{in}$, high energy high luminosity accretion produces
more outflow for both of the pseudo-potentials considered here. Also one 
can observe that for a fixed set of parameters 
$\left\{{\cal E}, {\dot M}_{Edd}, \gamma_{in}, \gamma_o \right\}$,
the amount of mass-outflow for $\Phi_2$ is higher compared to that 
in $\Phi_1$.\\
\noindent
In figure 3b, we plot $R_{\dot m}$ as a function of $T^{+}_{sh}$ for 
$\Phi_1$ (dotted line) and $\Phi_2$ (solid line) corresponding to the 
parameters used in the figure 3a.
We see that post-shock temperature non-linearly 
correlates with the energy of the
flow, and for a fixed accretion rate and adiabatic indices
of the inflow and outflow,
$R_{\dot m}$ also correlates with post-shock temperature, which indicates
that the outflow is thermally driven as well.

\subsection{Variation of $R_{\dot m}$ with adiabatic indices of the flow}

\noindent
In previous cases, the polytropic index $\gamma_{in}$ of the accreting matter
was always kept fixed at the value $\frac{4}{3}$. To have a better insight
of the behaviour of the outflow, we plot $R_{\dot m}$ as a function of
$\gamma_{in}$ (curves marked by $\gamma_{in}$ in Fig. 4) for a fixed value of ${\cal E}$ (= 0.00001)
 and
${\dot M}_{Edd}$ (= 1.0).
The upper 
range of $\gamma_{in}$ shown here is the range for which shock forms
for
the specified value of ${\cal E}$ and ${\dot M}_{Edd}$.
We have chosen the value of $\gamma_o$ in such a way so 
that the condition $\gamma_o<\gamma_{in}$ is always satisfied.
Defining $\Delta_\gamma$
to be 
${\Delta}_{\gamma}=\gamma_{in}-\gamma_o$,
we study the variation of ${R_{\dot m}}$ with $\gamma_{in}$ for 
${\Delta}_{\gamma} = 0.01$.
One can also obtain results for other values of ${\Delta}_{\gamma}$ in the same 
way.
We observe that ${R_{\dot m}}$ correlates with 
$\gamma_{in}$, which is expected because the specific enthalpy of the flow 
increases with $\gamma_{in}$ to produce a higher post-shock temperature for 
higher value of $\gamma_{in}$ (see also the curves marked by 
$T^{+}_{sh}$ in Fig. 4). We observe from our 
calculation that ${\rho}_{sh}^{+}$ and $u^{+}_{sh}$ correlates while $r_{sh}$ 
anticorrelates with $\gamma_{in}$. So increment of $\gamma_{in}$ satisfies 
all possible conditions to have a high value of $R_{\dot m}$.
In the same figure, curves marked by $T^{+}_{sh}$
show the variation of $R_{\dot m}$ with $T_{sh}^{+}$ corresponding to the 
values of $\gamma_{in}$ shown in the figure
(scaled 
$T^{+}_{sh}~\longrightarrow~T^{+}_{sh}\times10^{-11}$ to fit 
in the same figure) to show that 
here also the flow is thermally driven. 
Also, we observe that $M^{+}_{sh}$ as well as the shock compression ratio 
$R_{comp}$           
correlates with $\gamma_{in}$ so that `strong-shock' 
solutions are preferred to obtain a high value of mass-loss for this case.
So we conclude that as the accretion approaches from its ultra-relativistic nature
to its non-relativistic regime, the mass-loss rate increases.\\
\noindent
It is observed that for same set of parameters used to study the dependence 
of $R_{\dot m}$ on $\gamma_{in}$, flow in $\Phi_2$ produces more mass-loss
compared to flow in $\Phi_1$, i.e.,
$$
R_{\dot m}{\Bigg{\vert}}_{\Phi_2}~>~R_{\dot m}{\Bigg{\vert}}_{\Phi_1}
\eqno{(17)}
$$
for any value of $\gamma_{in}$ and the quantity
$$
{\delta}R_{\dot m}=\left[R_{\dot m}{\Bigg{\vert}}_{\Phi_2}-
R_{\dot m}{\Bigg{\vert}}_{\Phi_1}\right]
$$
decreases with decrease of $\gamma_{in}$. However, in the figure, the curves
showing the dependence of $R_{\dot M}$ with $\gamma_{in}$ are not properly
resolved to explicitly show the difference for flows in 
$\Phi_1$ and $\Phi_2$ due to the scaling used here to show 
both the temperature dependence of $R_{\dot M}$ as well as the dependence
of $R_{\dot M}$ on $\gamma_{in}$ in the same figure. \\ 
\noindent
We also study the variation of $R_{\dot M}$ with $\gamma_o$ (figure not
presented in this paper) for fixed values of $\left\{
{\cal E}, {\dot M}_{Edd}, \gamma_{in}\right\}$.
The general conclusion is that
$R_{\dot m}$ correlates with $\gamma_o$ for both of the pseudo-potentials.
This is because as $\gamma_o$
increases, shock location and post-shock density of matter does
not change (as $\gamma_o$ does not have any
role in shock formation or in determining the $R_{comp}$) but the sonic point
of the {\it outflow}
is pushed inward, hence  the velocity with which outflow leaves the 
shock surface goes up, resulting the increment in $R_{\dot m}$.
Here also we found that $R_{\dot m}$ for flow in $\Phi_2$ is always greater 
than than  $R_{\dot m}$ for $\Phi_1$ and the deviation becomes more 
prominent as the value $\gamma_o$ approaches to $\gamma_{in}$.
goes up, resulting the increment in $R_{\dot m}$.
%%%EKHANE

\section{Conclusion}
\noindent
In this paper, we could successfully construct a self-consistent spherically-symmetric,
polytropic, transonic, non-magnetized inflow-outflow system by simultaneously solving the
set of hydrodynamic equations governing the accretion and wind around a Schwarzschild black hole
using two different pseudo-potentials proposed by Artemova et. al.$^1$
Introducing a steady, standing, hadronic-pressure
supported spherical shock 
surface around the black hole as the effective physical atmosphere which may be responsible
for generation of accretion-powered spherical wind, we calculate the mass-outflow
rate 
$R_{\dot m}$ in terms of only three accretion parameters (conserved energy of the flow 
${\cal E}$, accretion rate ${\dot M}_{Edd}$
scaled in units of Eddington rate and polytropic index of the flow $\gamma_{in}$) and
only one outflow parameter (the polytropic index of the outflow, $\gamma_o$).
Not only do we provide a sufficiently plausible
estimation of $R_{\dot m}$,
we could also successfully study
 the dependence and variation
of this rate on various physical parameters governing the flow.\\
\noindent
The basic conclusions of this paper may be summarized as:
\begin{enumerate}
\item Shock formation is not a generic phenomena, i.e., not all solutions contain
shock, rather a specific region of parameter space spanned by ${\cal E}, {\dot M}_{Edd}$
and $\gamma_{in}$ allows shock formation. 
For given values of 
${\cal E}, {\dot M}_{Edd}$ and $\gamma_{in}$, while 
the value of shock location (in units of $r_g$)
{\it correlates} with 
${\dot M}_{Edd}$, it
{\it anti-correlates} with both ${\cal E}$ and $\gamma_{in}$.
High energy high luminosity purely non-relativistic accretion is a proper choice
to produce high post-shock proton temperature( thus to produce the maximum outflow). 
For other accretion parameters being the same, $\Phi_1$ produces a shock 
closer to the black hole compared to $\Phi_2$. This deviation is more prominent for
high ${\cal E}$ and $\gamma_{in}$.
\item The shock surface can serve as the `effective' physical barrier around the 
black hole regarding generation of mass loss via transonic spherical wind.
The fraction of accreting material being blown as wind (which is denoted as $R_{\dot m}$)
could be computed in terms of three accretion parameters and one outflow parameter.
\item $R_{\dot m}$ correlates with ${\cal E}, {\dot M}_{Edd}, 
\gamma_{in}$ and $\gamma_o$, 
Outflow could be generated for {\it both} sub-Eddington 
as well as super-Eddington accretion.
For a fixed set of \\
\noindent
$\left\{{\cal E}, {\dot M}_{Edd}, \gamma_{in}, \gamma_o\right\}$, mass 
outflow rate $R_{\dot m}$ is higher for flows in $\Phi_2$ compared to flows in
$\Phi_1$, which indicates that the barionic content of the spherical wind
is inversely proportional to the spatial gradient on the pseudo-potential 
used to study the problem though the exact physical reason behind this is 
not quite clear.
\item If a shock forms, then whatever the initial flow conditions and whatever the nature
of dependence of $R_{\dot m}$ on any of the accretion/ shock/ outflow parameters,
$R_{\dot m}$ normally 
 correlates with post-shock flow temperature, which indicates that
outflow is strongly thermally driven; hotter flow always produces more winds.\\
\end{enumerate}
\noindent
At this point, it is worth mentioning that the hot and dense
shock surface around black holes, which is proposed here as the effective
physical barrier around compact objects regarding the mass outflow, may
be generated due to other physical effects as well for spherical 
accretion.$^{10,11,12,13,14}$
Another very important
approach launched recently was
to construct such an
`effective barrier' for non-spherical disc accretion
to introduce the concept of CENtrifugal pressure
supported BOundary Layers (CENBOL).
Treating the CENBOL as the effective atmosphere of the rotating flows around
compact objects
(which forms as a result of standing Rankine-Hugoniot shock or due to the
maximization of polytropic pressure of accreting material in absence of shock),
detailed
computation of the mass-outflow rate from the advective accretion
disks has been done, and dependence 
of this rate on various
accretion and shock parameters has been quantitatively studied
by constructing a
self-consistent disk-outflow system.$^{15,16}$ \\
Our calculations in this paper, being
simply founded, do not explicitly include various radiation losses and cooling
processes, combined effects of which may reduce the post-shock proton temperature
(which means the reduction of outflow temperature),
in reality could be lower than what we have obtained
here and the amount of outflow would be less than what is obtained in our 
calculation.
This deviation will be more important for systems
with high accretion rates.
Nevertheless, cases of low accretion rates discussed here would not be affected
that
much and our preliminary investigation shows that even if we incorporate various
 losses,
the overall profile of the various curves showing the dependence of $R_{\dot m}$
on different inflow parameters would be exactly
the same, only the numerical value of $R_{\dot m}$
in some cases (especially for high accretion) might decrease.\\
\noindent
Although in this work we have performed our calculation for a 10$M_{\odot}$ Schwarzschild
black hole, general flow characteristics will be unchanged for black hole of any
mass except the fact that the region of parameter space responsible for shock formation
will be shifted and the value of $R_{\dot m}$ will explicitly depend on the
mass of the black hole.\\[0.25cm]
\nonumsection{Acknowledgements}
\vspace*{-0.5pt}
\noindent
The author would like to thank Prof. I. Novikov for useful discussion which 
enabled him to have a better understanding about the physical nature
of the pseudo-potentials used in this work.\\ [0.25cm]
\vspace*{-0.5pt}
\nonumsection{References}

\eject
\newpage
\begin{center}
\underline{\large\bf Figure Captions}\\[1cm]
\end{center}
\noindent
{{\bf Fig. 1:}
Variation of shock location $r_{sh}$ (plotted in logarithmic unit
along Y axis) with the specific energy of the flow ${\cal E}$ (plotted along
X axis) for three values of accretion rate (scaled in the unit of
Eddington rate ${\dot M}_{Edd}$) 0.25, 1.0 and 1.75 respectively from bottom
to top. While the dependence for flows in $\Phi_1$ is shown by
dotted lines, flows in $\Phi_2$ is shown by solid lines; see text for
details.}\\ \\
\noindent
{{\bf Fig. 2:}
 Combined solution topologies for transonic accretion-outflow
system for flows in $\Phi_1$ (2a) and $\Phi_2$ (2b) for fixed values
(shown in the figure) of accretion and outflow parameters.
While Mach number is plotted along the Y axis, the distance (in units of $r_g$)
from the
event horizon of the accreting black hole
is plotted along the X axis in logarithmic scale.
Solid curve marked
with
AB represents the pre-shock transonic accretion while four dotted curves
represent the accretion-powered outflow branches. Solid vertical line BC
(marked by $r_{sh}$)
with double arrow stands for the shock transition and solid line marked by
CD stands for the `self-wind' branch. $P_s$ is the location of the
inflow sonic point. See text for details.}\\ \\
\noindent
{{\bf Fig. 3a:}
Variation of $R_{\dot m}$ with inflow specific energy
${\cal E}$ for different values of accretion rate ${\dot M}_{Edd}$
shown in the figure. Dotted curves correspond to flow in $\Phi_1$ and
solid curves correspond to flow in $\Phi_2$.}\\ \\
\noindent
{{\bf Fig. 3b:}
 Variation of $R_{\dot m}$ with post-shock proton temperature
$T^{+}_{sh}$ corresponding to values of ${\cal E}$ shown in Fig. 3a.}\\ \\
\noindent
{{\bf Fig. 4:}
Variation of $R_{\dot m}$ with the polytropic index of accretion
$\gamma_{in}$ (curves marked by $\gamma_{in}$) and the corresponding
post-shock temperature (curves marked by $T^{+}_{sh}$) for flows in $\Phi_1$
(dotted curve) and in $\Phi_2$ (solid curve). The Mass-outflow rate
non-linearly correlates with $\gamma_{in}$ as well as with post-shock
proton temperature which indicates that the outflow is thermally driven
as well, see text for details.}\\ 
\newpage
\begin{figure}
\vbox{
\vskip -4.5cm
\centerline{
\psfig{file=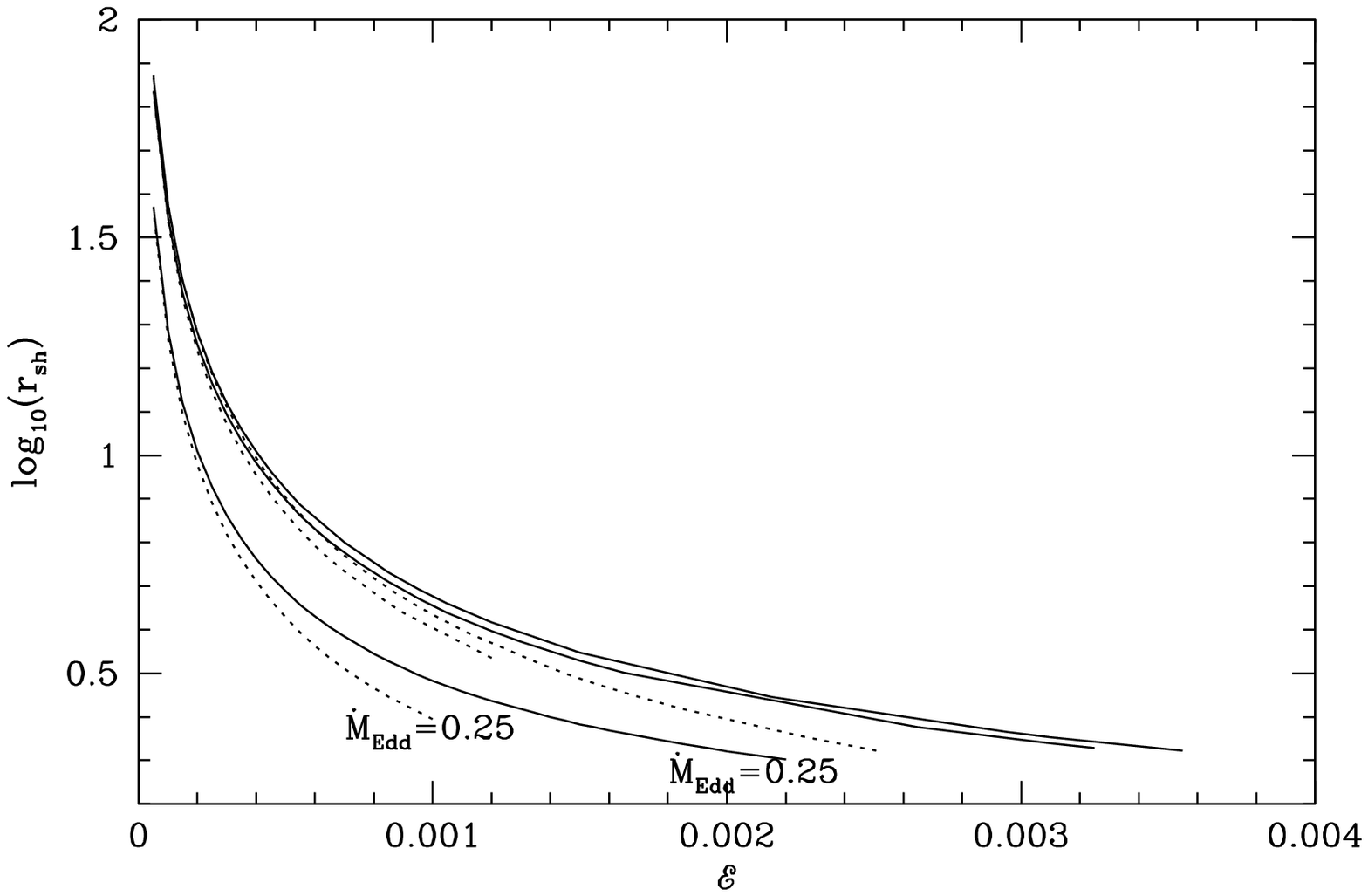,height=19cm,width=18cm,angle=0.0}}}
\noindent {{\bf Fig. 1:}}
\vskip 1.0cm
\noindent
{{\bf Author:} Tapas Kumar Das. \\
{\bf Titile of the paper:} Pseudo-Schwarzschild description of 
accretion-powered spherical outflow}
\end{figure}
\newpage
\begin{figure}
\vbox{
\vskip -4.5cm
\centerline{
\psfig{file=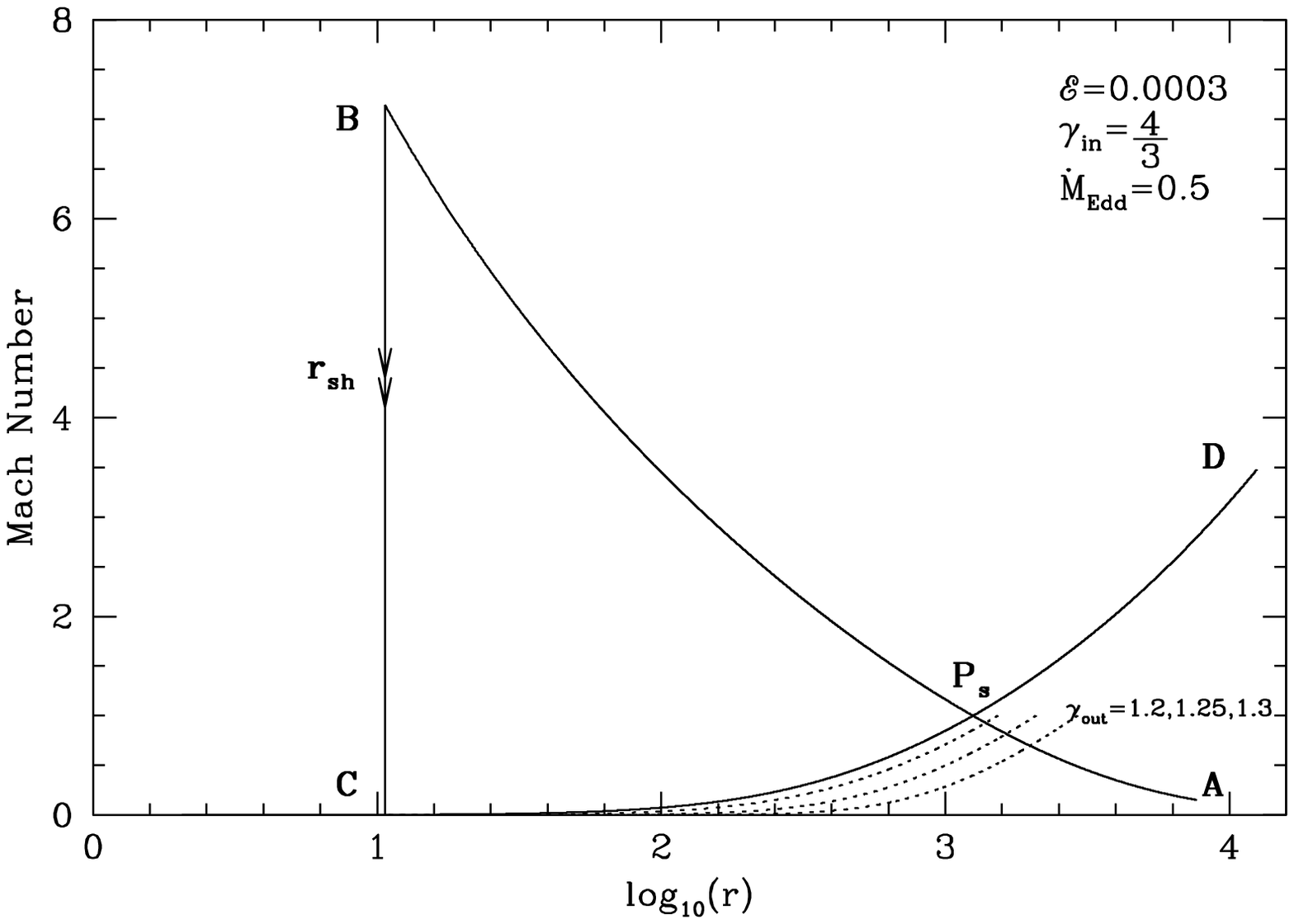,height=19cm,width=18cm,angle=0.0}}}
\noindent {{\bf Fig. 2a:}}
\vskip 1.0cm
\noindent
{{\bf Author:} Tapas Kumar Das. \\
{\bf Titile of the paper:} Pseudo-Schwarzschild description of 
accretion-powered spherical outflow}
\end{figure}
\newpage
\begin{figure}
\vbox{
\vskip -4.5cm
\centerline{
\psfig{file=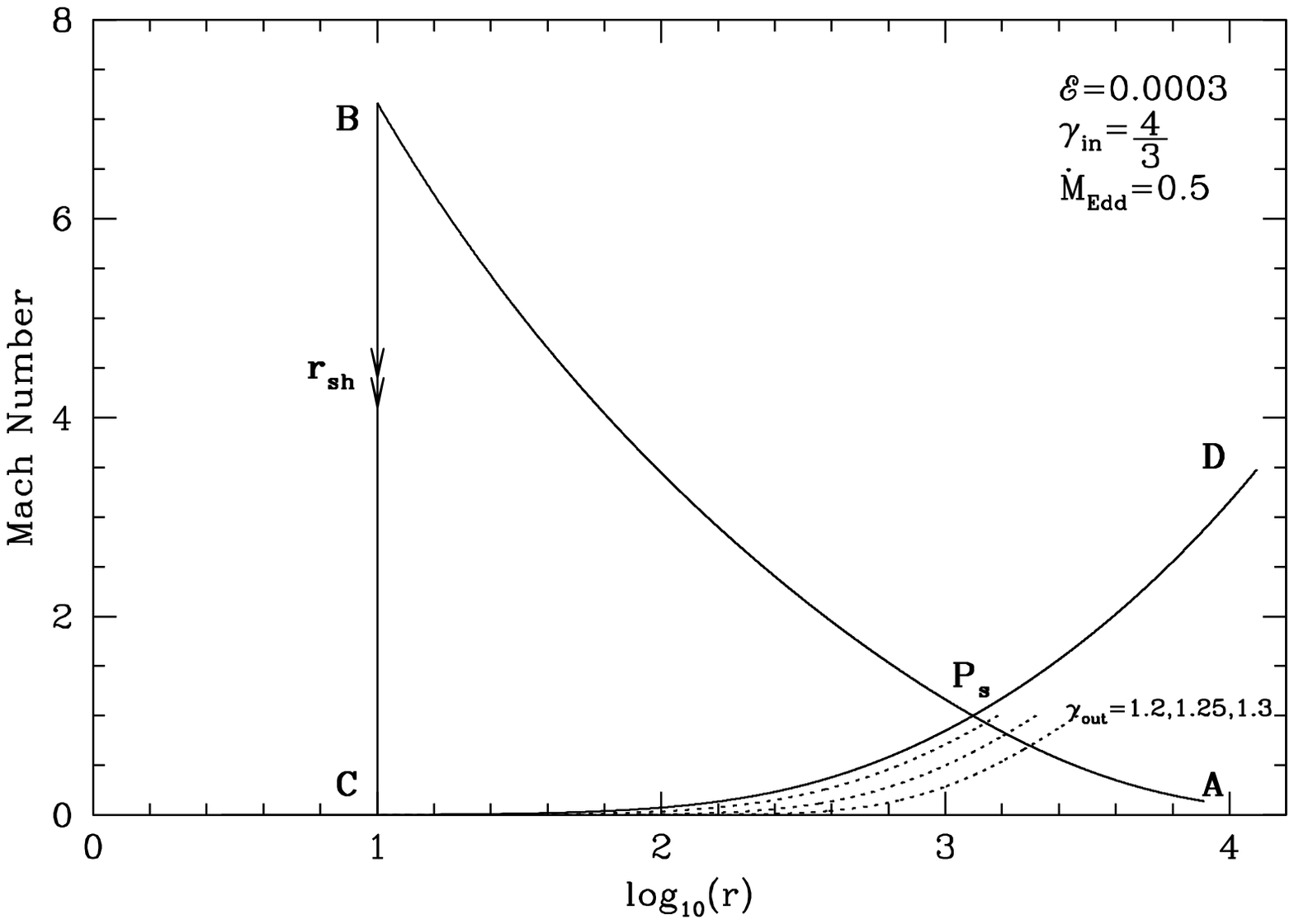,height=19cm,width=18cm,angle=0.0}}}
\noindent {{\bf Fig. 2b:}}
\vskip 1.0cm
\noindent
{{\bf Author:} Tapas Kumar Das. \\
{\bf Titile of the paper:} Pseudo-Schwarzschild description of 
accretion-powered spherical outflow}
\end{figure}
\newpage
\begin{figure}
\vbox{
\vskip -4.5cm
\centerline{
\psfig{file=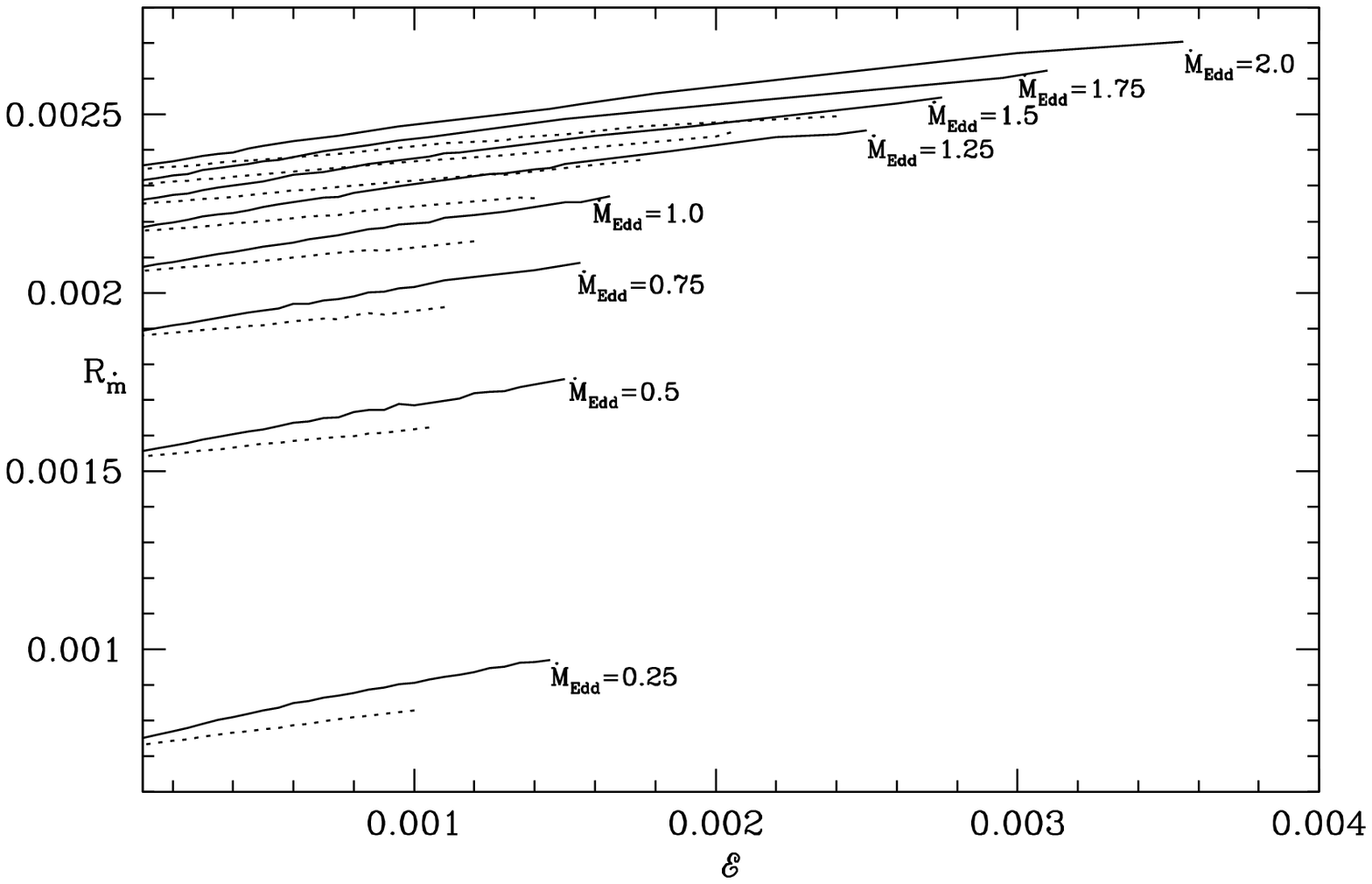,height=19cm,width=18cm,angle=0.0}}}
\noindent {{\bf Fig. 3a:}}
\vskip 1.0cm
\noindent
{{\bf Author:} Tapas Kumar Das. \\
{\bf Titile of the paper:} Pseudo-Schwarzschild description of 
accretion-powered spherical outflow}
\end{figure}
\newpage
\begin{figure}
\vbox{
\vskip -4.5cm
\centerline{
\psfig{file=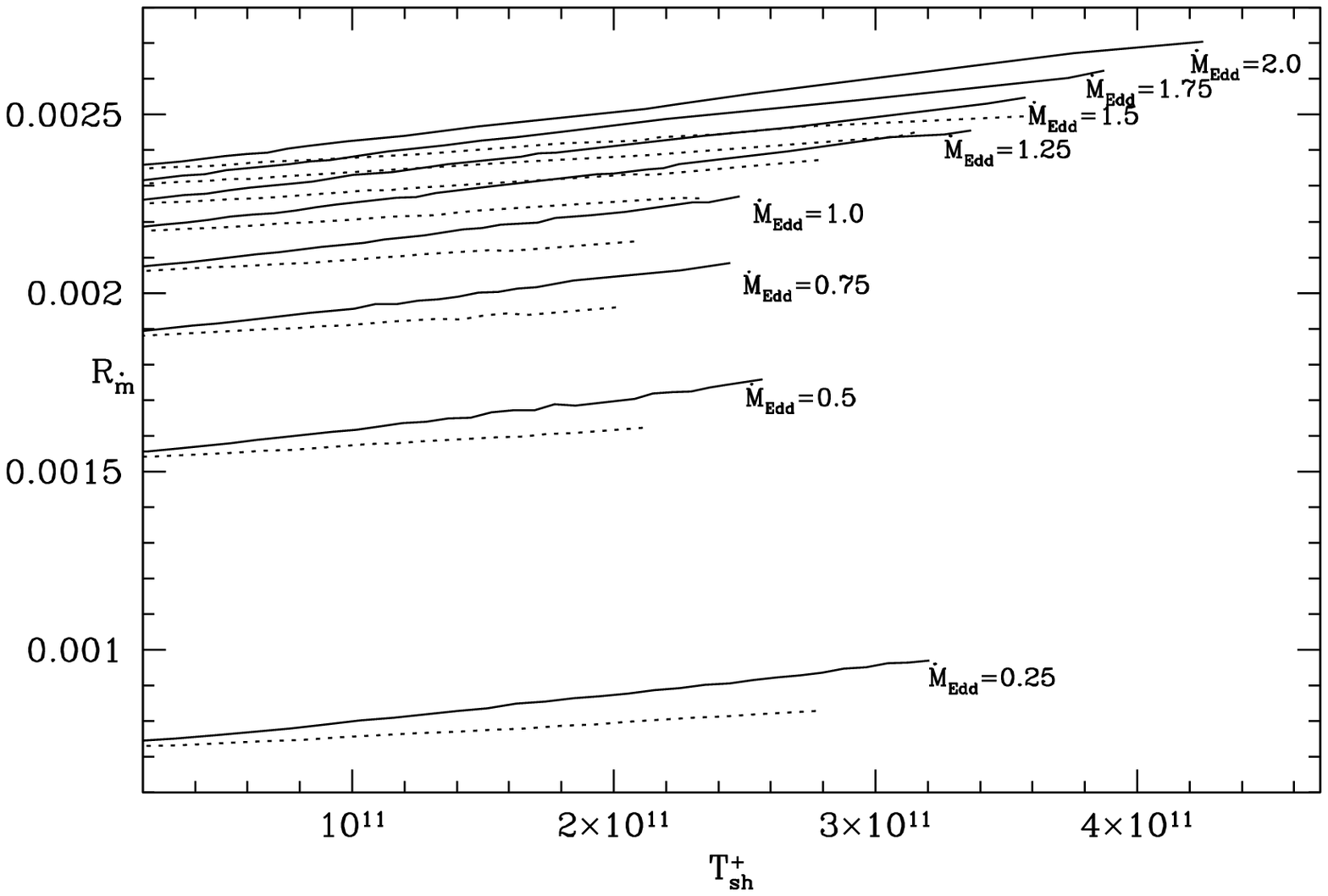,height=19cm,width=18cm,angle=0.0}}}
\noindent {{\bf Fig. 3b:}}
\vskip 1.0cm
\noindent
{{\bf Author:} Tapas Kumar Das. \\
{\bf Titile of the paper:} Pseudo-Schwarzschild description of 
accretion-powered spherical outflow}
\end{figure}
\newpage
\begin{figure}
\vbox{
\vskip -4.5cm
\centerline{
\psfig{file=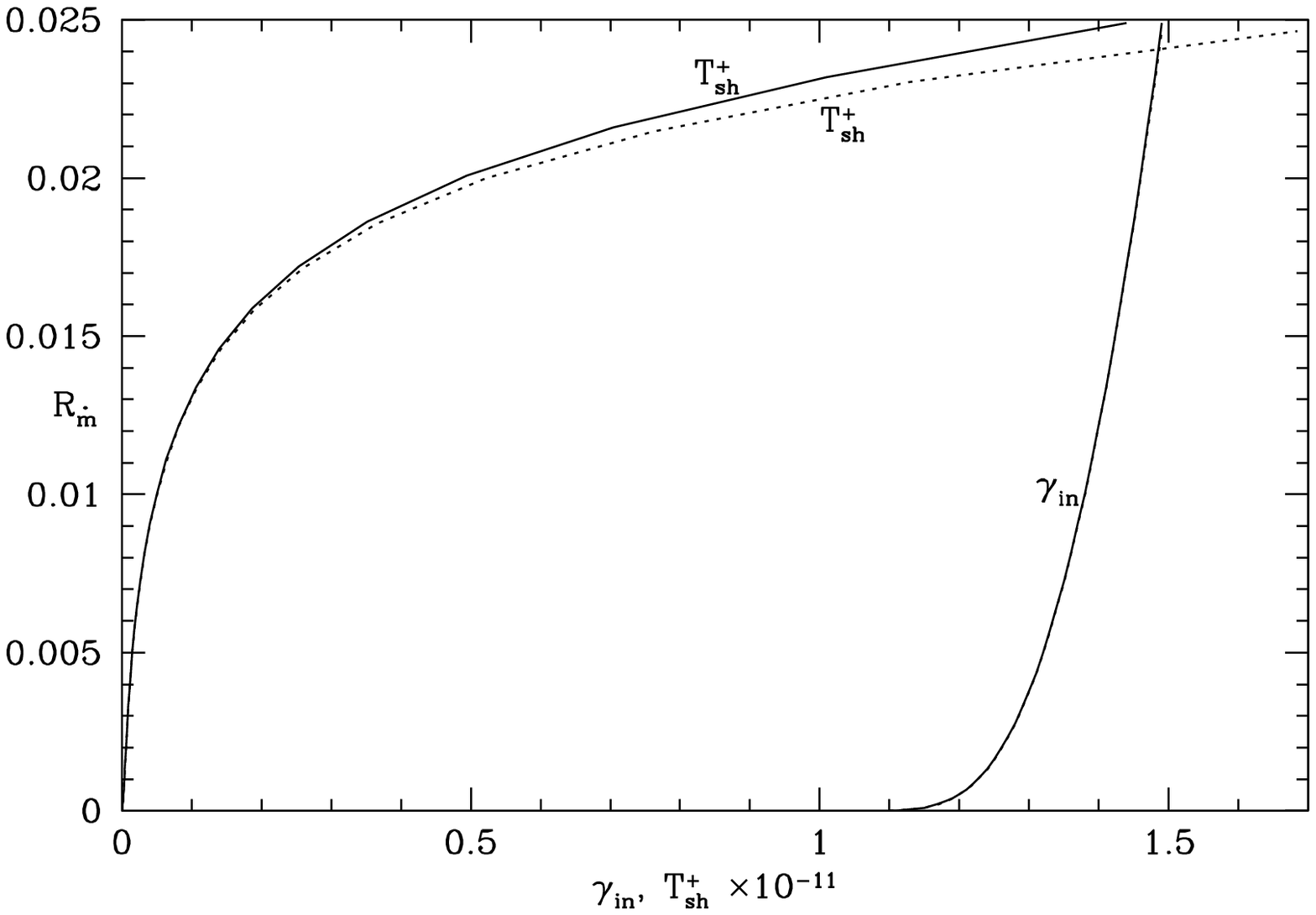,height=19cm,width=18cm,angle=0.0}}}
\noindent {{\bf Fig. 4:}}
\vskip 1.0cm
\noindent
{{\bf Author:} Tapas Kumar Das. \\
{\bf Titile of the paper:} Pseudo-Schwarzschild description of 
accretion-powered spherical outflow}
\end{figure}
\newpage

\end{document}